# Demystifying Cryptocurrency Mining Attacks: A Semi-supervised Learning Approach Based on Digital Forensics and Dynamic Network Characteristics

Aaron Zimba[1], Mumbi Chishimba[2], Christabel Ngongola-Reinke[3], Tozgani Fainess Mbale[4]

[1]Department of Computer Science & IT, Mulungushi University, Kabwe, Zambia, [2]Department of Information Technology National Institute of Public Administration (NIPA), Lusaka, Zambia, [3]Department of Economics, Mulungushi University, Kabwe, Zambia, [4]Department of Electrical & Electronic Engineering, University of Zambia, Lusaka, Zambia

[1]gvsfif@gmail.com, [2]chishimba.mumbi@gmail.com, [3]christabel.ngongola@gmail.com, [4]tozganimbale@gmail.com

*Abstract*—Cryptocurrencies have emerged as a new form of digital money that has not escaped the eyes of cyber-attackers. Traditionally, they have been maliciously used as a medium of exchange for proceeds of crime in the cyber dark-market by cyber-criminals. However, cyber-criminals have devised an exploitative technique of directly acquiring cryptocurrencies from benign users' CPUs without their knowledge through a process called crypto mining. The presence of crypto mining activities in a network is often an indicator of compromise of illegal usage of network resources for crypto mining purposes. Crypto mining has had a financial toll on victims such as corporate networks and individual home users. This paper addresses the detection of crypto mining attacks in a generic network environment using dynamic network characteristics. It tackles an in-depth overview of crypto mining operational details and proposes a semi-supervised machine learning approach to detection using various crypto mining features derived from complex network characteristics. The results demonstrate that the integration of semi-supervised learning with complex network theory modeling is effective at detecting crypto mining activities in a network environment. Such an approach is helpful during security mitigation by network security administrators and law enforcement agencies.

## I. INTRODUCTION

The general aim of conventional cyberattacks has generally been to obtain monetary proceeds of the associated cybercrime. Attackers have had the challenge of acquiring these monetary proceeds with little or no monetary trail since conventional payments leave a trail of traceable financial activities [1]. Cryptocurrencies have alleviated this challenge as they provide for privacy and anonymity [2]. The strong privacy provided in cryptocurrencies makes it almost impossible to trace financial payments [3]. As such, cryptocurrencies have become a de facto method of payments in most finance-related cyber-attacks [4], a trend not uncommon in crypto-ransomware attacks.

However, since victims of cybercrime have had the ability to make payments in cryptocurrencies such as Bitcoin (implying users store cryptocurrencies on computers), attackers have now moved on to attack the very user cryptocurrencies from digital wallets as was evidenced in various attacks [5]. Furthermore, it is not uncommon to find financial malware that seeks to steal cryptocurrencies from targeted users as an extra functionality [6]. Since not all targeted users harbor cryptocurrencies, attackers have devised a technique of directly generating cryptocurrencies from the victims' CPU (crypto mining[1]) by enlisting them to a mining pool. The cryptocurrencies are generated by installable malware or via browser-based crypto mining. Victims are enlisted in a crypto mining pool since solo mining is not efficient [7]. As such, corporate or enterprise networks are attractive to crypto mining attackers because they provide a pool of devices for crypto mining. It is thus not uncommon to find illegal crypto mining cloud computing and IoT environments [8] as well as critical infrastructure systems such as SCADA [9]. Illegal crypto-mining has since been on the rise and costed victims millions of dollars [10]. Consequently, the year 2018 saw the growth of crypto mining malware by 4,000% [11]. As such, crypto mining attacks have proven to be a force to reckon with which can longer be avoided even as attackers have been eschewing the infamous ransomware attacks [12]. The diagram in Figure 1 shows the decline in ransomware attacks versus the rise in crypto mining attacks according to the IBM-X-Force Threat Intelligence Index 2019 [13].

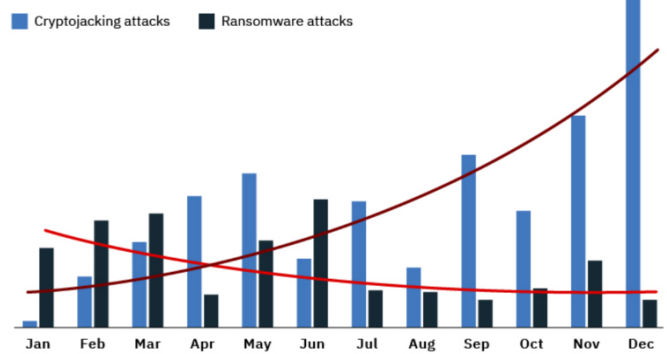

Fig. 1. Crypto mining vs Ransomware attacks in 2018 [13]

Crypto mining is taking over ransomware owing to its ease of administration; easily proliferated by phishing emails, no user input required and the difficulty associated with tracing the perpetrators. These advantages have seen an increase in the stealthier attacks, i.e. crypto mining, by 450% even as cybercriminals pivot from the common ransomware attacks [14]. Crypto mining attacks present a million-dollar industry [15].

In Africa, crypto mining is particularly prevalent in Ethiopia, Tanzania, and Zambia which account for 3 of the top-5 countries largely impacted by crypto-mining attacks, according to a Microsoft report [16]. The most impacted victims are SMEs as the security therein is not as robust as in larger corporations.

---
[1] Crypto mining is a process of using the resources of a computer system to mine cryptocurrencies.

Like all malware activities and cyberattacks, crypto mining activities generate noise in the form of network traffic. However, the types of network characteristics associated with these types of attacks are peculiar to crypto mining in that victims enlisted to a mining pool or botnet needs to communicate with the associated C2 servers and mining servers. As such, the detection of crypto mining activities in a network environment calls for an approach that takes into consideration these network characteristics. This paper addresses the detection of crypto mining attacks in a generic network environment using dynamic network characteristics. It tackles an in-depth overview of crypto mining operational details and proposes a machine learning approach to detection using various crypto mining features derived from the network characteristics. The Small-World network models [17] of complex network evolution theory are adopted for attack modeling and we use a semi-supervised approach to machine learning for detection.

The rest of the paper is organized as follows; Section II presents the related works while the methodology and proposed detection framework are brought forth in Section III. The results and the analyses thereof real-world in Section IV and the conclusion is drawn in Section V.

## II. RELATED WORKS

Even though crypto-mining attacks are a fairly new phenomenon, they have attracted significant attention in the security landscape. Some research works have concentrated on crypto mining in general computer systems [18] whilst others have narrowed the scope to critical infrastructure and IoT [19]. Authors in [20] propose an end-to-end analysis of browser-based crypto mining by statically and dynamically examining the rise of crypto mining in the real world cases. The proposed approach inspects the traversing traffic between web-sockets without blacklisting of IP addresses. They achieve a detection accuracy of 96.4% using code analysis.

Authors in [21] propose a host-based approach to crypto mining called BotcoinTrap. Modeling via dynamic analysis of executable binary crypto mining files to detect Bitcoin-mining botnets is adopted. The advantage of this approach is that it can detect Bitcoin mining botnets at the lowest level of execution. The Bitcoin block header is centrally used as the pivotal piece of information in this detection methodology. The drawback of this approach is that it specifically applies only to the detection of Bitcoin miners, whereas the crypto mining landscape has seen the emerging of competing and easy-to-mine cryptocurrencies such a Monero and Ethereum.

In [22], the authors examine recent trends towards in-browser mining of cryptocurrencies. They concentrate their efforts on the mining of Monero cryptocurrency via CoinHive and those of similar code-bases. In their model, a web user visits a vulnerable site infected with JavaScript code that executes on the client-side browser, thus mining a cryptocurrency without the user's consent. They further survey the crypto mining landscape in order to conduct measurements to establish the prevalence and profitability thereof. They outline the ethical framework for classifying the attack as an inherent attack or business opportunity. They delineate the various stages involved in the process crypto mining process and thereafter brief the various terms associated with crypto mining. However, their approach does not address the systematic detection of crypto mining.

In [23], the authors approach crypto mining detection using dynamic opcode analysis on non-executable files. They use a specified dataset to achieve high detection rates of browser-based crypto mining using Random forest (RF) as the preferred classification algorithm. Their model distinguishes between crypto mining websites, weaponized benign crypto mining websites, de-weaponized crypto mining websites, and real-world benign crypto mining websites. As such, their technique offers an opportunity not only to detect but to prevent as well as mitigate crypto-mining attacks.

Authors in [24] present an in-depth analysis of the crypto mining operation. They designed and implemented a passive-active flow monitoring and catalog to detect crypto-mining activities from compromised devices in a network. They tested the feasibility of their approaches to real-life data where passive-active detection is capable of discovering emerging or deliberately hidden crypto mining pools.

## III. METHODOLOGY AND PROPOSED FRAMEWORK

The enlisting of vulnerable and exploitable devices to a crypto mining pool is a dynamic process that can be viewed as an evolution of node-addition or deletion in an attack graph. Since the nodes in the mining pool interact one with the other and with the central server, the system can thus be characterized by vertex degrees and clustering coefficients. It is on this premise that we employ the use of Small-World network models of complex network theory to depict the behaviour of the attack network and deduce the corresponding features for purposes of detection. The diagram in Figure 2 shows a time-slice crypto-mining depicting the initialization and growth of a crypto mining pool in a target network.

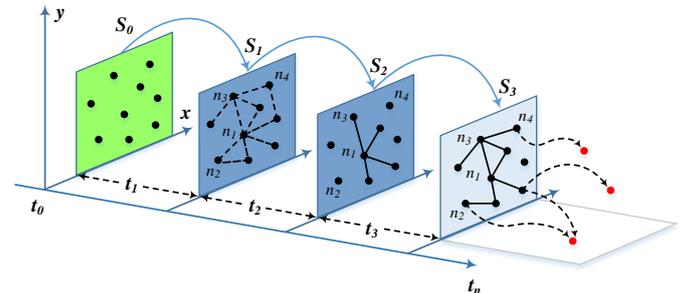

Fig. 2. Dynamic growth of crypto mining pool at various stages of the crypto mining process

The first phase $S_0$ represents the state of the targeted network system before any device is listed to a crypto mining pool. At this phase, the devices in the network are only susceptible to crypto mining but not yet compromised. As such, the vertex degree and clustering are equal to zero. In phase $S_1$, vulnerable nodes are identified and consequently added to the crypto mining pool in phase $S_2$. It is worth noting that at this stage, the vertex degree and clustering coefficients are now greater than zero.

$$S_n(t) \rightarrow S_{n+1}(t+1) := \begin{cases} (N_t \cup n_{t+1}^+) - (n_{t+1}^-) \\ \cdot \\ (E_t \cup e_{t+1}^+) - (e_{t+1}^-) \end{cases} \quad (1)$$

Phase $S_3$ represents active crypto mining where the enlisted victims in the crypto mining pool are coordinating and working together towards the associated proof-of-work[2]. Equation (1) depicts the dynamic transitions of a victim device enlisted to a mining pool at a point in time as echoed in Figure 2.

---
[2] Proof-of-work refers to the cryptographic computational puzzle that miners have to solve in order to be issued a crypto currency unit.

As the state of the enlisted victim devices transitions from state $S_0$ to $S_n$, a series of network traffic is generated which we use to derived features for the detection process.

Members enlisted in a crypto mining pool used dedicated protocols to coordinate the distributed mining process. The 3 common TCP-based crypto mining protocols are GetWork, GetBlockTemplate, and Stratum protocol [25]. Other traffic details found in crypto mining pools include registration and authentication traffic, recurrent assignment of work packages provided by the crypto mining server. It is from these dynamic traffic details that we draw features to devise a detection methodology. In light of this, we present a semi-supervised learning approach to crypto mining detection that takes advantage of the huge amount of unclassified dataset [26] to perform classification of suspicious hosts participating in crypto mining activities and using few labeled instances from the labeled data. The proposed detection framework is shown in Figure 3.

ability to handle clusters of varying sizes, densities and shapes. As such, two nodes that are relatively close but belong to different clusters are handled effectively.

Algorithm 1 illustrates the enhanced SNN algorithm. As shown from Figure 3, our semi-supervised approach consists of two phases: 1) an unsupervised phase that produces complex network characteristics features based on vertex degrees and clustering coefficients. 2) a supervised phase that learns and trains the model. This phase uses the KNN classifier and the labeled data. In short, our semi-supervised learning approach uses the unsupervised learning method to extract features from the unlabeled dataset and the supervised model classifies this data instances of crypto mining using complex network characteristics features. The unsupervised phase utilizes the shared nearest neighbor clustering whilst the supervised phase utilizes the KNN.

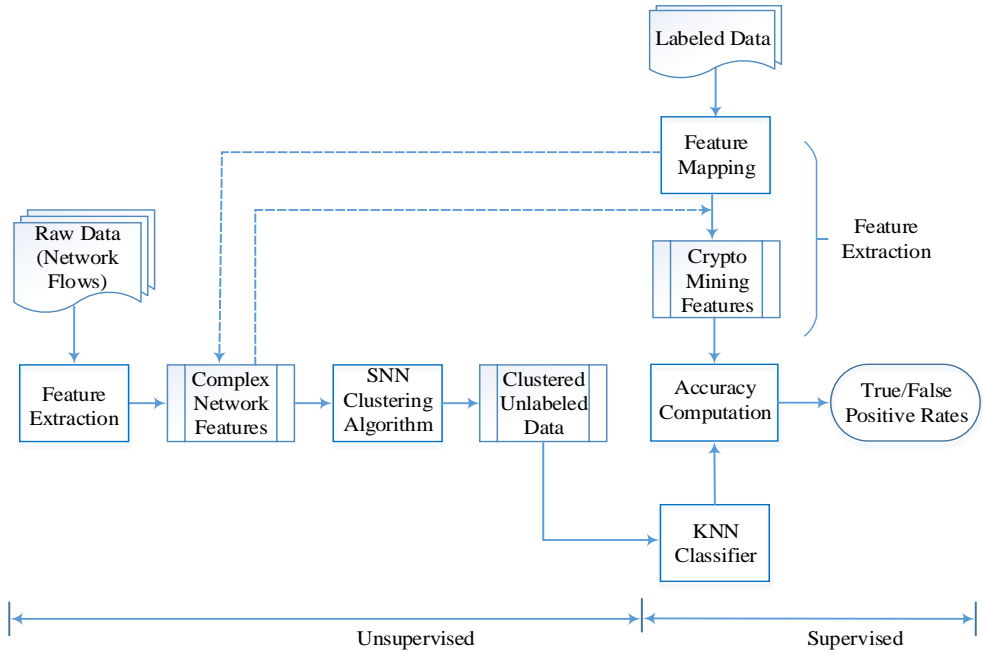

Fig. 3. The semi-supervised approach to crypto mining detection

Our semi-supervised approach shown in Figure 3 shows that in addition to unlabeled raw data (network flow traffic), we have a set of labeled data with features depicted in Table 1. Our semi-supervised approach uses complex network characteristics features of unlabeled data (clustering coefficients and vertex degrees) to create a supervised model. The feature extraction step in the supervised section of our approach uses a mapping scheme to extract hosts from the unlabeled dataset.

We propose a semi-supervised learning approach where we first derive different clusters mainly based on the clustering coefficient and vertex degree. To analyze the normalized data and detect crypto mining, we employ an enhanced semi-supervised algorithm based on the Shared Nearest Neighbour (SNN) clustering algorithm [27]. The SNN clustering defines similarity or proximity between two nodes in terms of the number of directly connected neighbors they have in common. This suits its applicability in complex networks since the clustering coefficient and vertex degrees are dictated by neighbor relations. As such, we adopt the SNN algorithm which apart from considering direct associations between nodes also considers indirect connections. This provides for an ability to detect similarities between nodes that are not necessarily adjacent. Additionally, SNN has the

| Algorithm 1: Enhanced SNN for detecting crypto mining |
|---|
| **Input**: $G$ -undirected graph, $k$ number of shared nearest neighbors |
| **Output**: $L^*$ - list of suspicious hosts participating in crypto mining. |
| 1 Initialize $G^*$ with $|V(G)|$ vertices, no edges |
| 2 **foreach** $i = 1\ to\ V(G)$ **do** |
| 3    **foreach** $j = i + 1\ to\ V(G)$ **do** |
| 4      $counter = 0$ |
| 5      **foreach** $m = 1\ to\ V(G)$ **do** |
| 6       **if** vertex $i$ and vertex $j$ both have an edge with vertex $m$ **then** |
| 7         $counter = counter + 1$ |
| 8      **end** |
| 9      **if** $counter \geq k$ **then** |
| 10        Connect an edge between vertex $i$ and vertex $j$ in $G^*$ |
| 11        **for** $S \leftarrow 0$ |
| 12          **if** $\triangle K_i > 1$ at time $t^*$ for external communications |
| 13            **then** $(H_{int-src}) \in S_1$ |
| 14        **else if** $\triangle K_i'' > \triangle K_i\ \&\&\ \triangle C_i > 1\ \&\&\ \triangle C_i > [\triangle C_{i-1,\ldots 0}]$ |
| 15            **then** $(H_{int-src}) \in S_2$ |
| 16        **else if** $\triangle K_i'' > 1\ \&\&\ M_v > (X^*_{threshold})$ for time window $\triangle t$ |
| 17            **then** $(H_{int-src}) \in S_3$ |
| 18      **end if** |
| 19    **end** |
| 20 **end** |
| 21 **Return** $L^*$ |

The semi-supervised learning approach is summarized in Algorithm 2.

Algorithm 2: Semi-supervised learning for crypto mining detection

Input: $X_{u-d}^i = \{x_{l-d}^1, x_{l-d}^2, x_{l-d}^3, \ldots, x_{l-d}^i\}$, unlabeled network flows

where $x_{l-d}^i \in R^n, i = 1,2,3,\ldots,n$

: $X_{l-d}^i = \{x_{u-d}^1, x_{u-d}^2, x_{u-d}^3, \ldots, x_{u-d}^i\}$, labeled data

where $x_{u-d}^i \in R^n, i = 1,2,3,\ldots,n$

Output: $TP$ && $FP$ rates – cryptocurrency mining detection accuracy

1. Read the unlabeled & labeled network flow dataset
2. Normalize original data $X_{l-d}^i$, $X_{u-d}^i$, to get the data $\overline{X_{u-d}^i}$, $\overline{X_{l-d}^i}$
3. Extract $k$ and $c$, complex network features from $X_{u-d}^i$
4. From $X_{l-d}^i$, map corresponding $k$ and $c$ in $X_{u-d}^i$, generate features
5. Cluster $X_{u-d}^i \rightarrow SNN \Rightarrow C_0, C_1, C_2, \ldots, i$
6. Generate cluster states $C_0, C_1, C_2, \ldots, i \rightarrow FSM\ (S_n) \Rightarrow C_i^{S_n}$
7. Train supervised KNN with $\overline{X_{l-d}^i}$ and $\overline{X_{u-d}^i}$
8. Classify $C_i^{S_n}$ with the KNN
9. Compute detection accuracy $TP$ & $FP$ rates based on SNN & KNN
10. Return $TP$ && $FP$ rates for $C_i^{S_n}$ clusters

The unlabeled and labeled data $\{X_{u-d}^i, X_{l-d}^i\}$ from network flows and the labeled data respectively are initialized and read in step 1. In step 2, we normalize the data by converting the input values to a common scale $[\overline{X_{u-d}^i}, \overline{X_{l-d}^i}]$. This enables us to make an effective comparison of the variations of the clustering coefficient and vertex degree. The complex network characteristics feature $k$ and $c$ are extracted from the unlabeled data $\{X_{u-d}^i\}$ in step 3. The labeled data is used in a mapping scheme in step 4 to locate a host and generate crypto mining features from the dataset. The SNN unsupervised clustering algorithm is used to create clusters in step 5 via the process $\{X_{u-d}^i\} \rightarrow SNN \Rightarrow C_0, C_1, C_2, \ldots, i$. Step 6 assigns states to the clusters $C_0, C_1, C_2, \ldots, i$ generated in step 5. The KNN supervised algorithm is applied to the datasets $\{\overline{X_{l-d}^i}\}$ and $\{\overline{X_{u-d}^i}\}$ in step 7. The clustered data with states $\{C_i^{S_n}\}$ is classified by the KNN algorithm in step 8. The True Positive and False Positive rates ($TP$ && $FP$) are computed in step 9 and the corresponding $TP$ && $FP$ rates for the clustered states are returned in step 10.

## IV. RESULTS ANALYSIS AND DISCUSSIONS

To apply the aforementioned framework and algorithms, we first start by analyzing the unlabeled traffic content with a protocol analyzer for crypto mining and non-crypto mining TCP and UDP traffic. The results are shown in Figure 4 below.

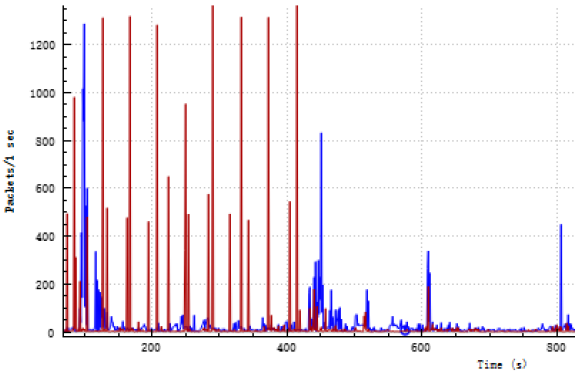

Figure 4. TCP (blue) and UDP (red) traffic from the dataset

We analyze traffic with crypto mining activities for the cryptocurrencies Ethereum, Monero, and Zcash for their corresponding mining pools. The diagram in Figure 5 shows traffic for the Ethereum mining pool captured via Wireshark.

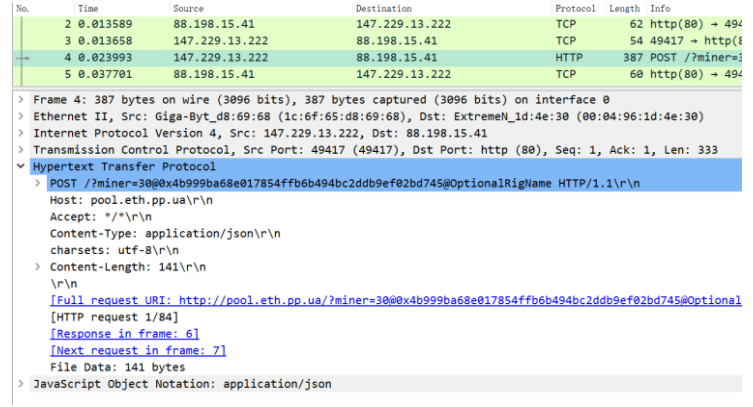

Fig. 5. Traffic for the Ethereum mining pool

It was noted that the mining protocols leverage TCP as the transport layer protocol. In comparison to official crypto p2p clients, the mining protocols do not necessarily use "well-known" port numbers. This is all dependent on the configuration of the administrator. As such, it is not uncommon to encounter port numbers for http=80, https/TLS=443, and SMTP=25 being used to bypass firewalls between the mining victim and the corresponding mining pool server.

The feature vector specified in this dataset [26] which we later use for the supervised learning stage is outlined in Table I.

TABLE I. DATASET FEATURE VECTOR

| SN | Feature | Description |
|---|---|---|
| 1 | bpp | Bytes per packet per flow per all flows |
| 2 | ppm | Packets per minute |
| 3 | ppf | Packets per flow per all flows |
| 4 | Ackpush_all | Number of flows with ACK+PUSH flags to all flows |
| 5 | Req_all | Request flows to all flows |
| 6 | Syn_all | Number of flows with SYN flag to all flows |
| 7 | Rst_all | Number of flows with RST flag to all flows |
| 8 | Fin_all | Number of flows with FIN flag to all flows |
| 9 | class | class - miner or not-miner |

We apply different clustering algorithms in Weka [28] and later compare them with the results of SNN clustering. The table below Table II shows the results of different clustering algorithms.

TABLE II. COMPARISON RESULTS OF UNSUPERVISED LEARNING

| SN | Clustering Model | Clustered Instances | Distribution (%) |
|---|---|---|---|
| 1 | Simple-K-Means | 2 [C0, C1] | 55% : 45% |
| 2 | Canopy | 3 [C0, C1, C2] | 55% : 4% : 41% |
| 3 | MakeDensityBased Clusterer | 2 [C0, C1] | 55% : 45% |
| 4 | HierarchicalClusterer | 2 [C0, C1] | 0% : 100% |
| 5 | FilteredClusterer | 2 [C0, C1] | 55% : 45% |
| 6 | FarthestFirst | 2 [C0, C1] | 100% : 0% |
| 7 | SNN | 5 [C0, C1, C2, C3, C4] | 22%:19%:18%:15%:26% |

Application of the SNN clustering algorithm produces 5 clusters of different properties. Table II shows 5 clusters with IDs C0, C1, C2, C3, and C4. As can be seen from Table II, the SNN algorithm performs better clustering with not only the highest numbers of clusters but even a better distribution.

Cluster C0 has a high *bpp* (97.3%) and a high *ppm* (79.2%). It also has a high *ppf* (65.8%) compared to *Ackpush_all (47.1%)*. This implies that hosts in this cluster have a higher vertex degree and clustering coefficient with regards to external communications.

On the contrary, cluster C2 has *bpp (98.5%)* and *ppm (83.6%)* but the *Ackpush_all (90.6%)* is greater than *ppf (75.3%)*. This implies that hosts in this cluster have a higher clustering coefficient and vertex degree with regards to internal communications. The number of flows with FIN flags to all flows for activities in this cluster is relatively higher than C0.

A lower *bpp (1.6%)* and a high *ppm (99.6%)* corresponding to *ppf (33.1%)* instead of *Ackpush_all (1.8%)* for a smaller time window in cluster C4 entail that hosts in this cluster communicate more with external hosts. Furthermore, hosts in this cluster have a high *Syn_all (97.8%)* value implying a high number of synchronization connections requests to the mining pool.

TABLE III. CLUSTERING RESULTS

| Attribute | Cluster-ID | | | | |
|---|---|---|---|---|---|
| | *C0* | *C1* | *C2* | *C3* | *C4* |
| *bpp* | 0.973 | 0.763 | 0.985 | 0.861 | 0.016 |
| *ppm* | 0.792 | 0.762 | 0.836 | 0.582 | 0.996 |
| *ppf* | 0.658 | 0.371 | 0.753 | 0.864 | 0.331 |
| *Ackpush_all* | 0.471 | 0.937 | 0.906 | 0.743 | 0.018 |
| *Req_all* | 0.984 | 0.735 | 0.969 | 0.791 | 0.092 |
| *Syn_all* | 0.548 | 0.524 | 0.81 | 0.577 | 0.978 |
| *Rst_all* | 0.471 | 0.832 | 0.988 | 0.72 | 0.511 |
| *Fin_all* | 0.35 | 0.526 | 0.871 | 0.936 | 0.302 |

The clusters C1 and C3 have relatively average network statistics that depict the behaviour of benign hosts. The high *Ackpush_all (93.7%)* in C1 corresponds to a high *Rst_all (83.2%)* which is a correlation expected of normal network traffic. Equally in cluster C3, *bpp (86.1%)* corresponding to *ppf (86.4%)* which is supplemented by average values of other characteristics in the same range. The variations in the clustering coefficient and vertex degree in these network traffic statistics in the respective clusters depict the overall movement of the movement vector from the feature centroid.

After generating the clusters and associating them with crypto mining instances, we use the labeled dataset for classification and evaluate the effectiveness of our proposed approach. This is because the hosts in the labeled data are technically labeled as malicious for generating crypto mining traffic. However, we do not evaluate which stage of the crypto mining process the traffic belongs to. The detailed characteristics of the model for the hosts classified using the clusters, which are the results of the classification are shown in Table IV.

TABLE IV. MODEL CHARACTERISTICS CRYPTO MINING DETECTION

| Class | TP Rate | FP Rate | Precision | Recall | F Measure | MCC | ROC Area | PRC Area |
|---|---|---|---|---|---|---|---|---|
| Not Miner | 0.998 | 0.462 | 1 | 0.998 | 0.999 | 0.276 | 0.974 | 1 |
| Miner | 0.538 | 0.002 | 0.143 | 0.538 | 0.226 | 0.276 | 0.974 | 0.504 |
| Avg. | 0.997 | 0.461 | 0.999 | 0.997 | 0.998 | 0.276 | 0.974 | 1 |

Correctly classified instances represent 99.72% while incorrectly classified instances represent 0.28%. The diagram in Figure 6 shows the confusion matrix of the correctly and wrongly classified instances.

| Classified as | Was actually | |
|---|---|---|
| | **Not Miner** | **Miner** |
| **Not Miner** | 355692 | 882 |
| **Miner** | 26 | 147 |

Figure 6. Confusion matrix for the model

The model has good performance because the weighted average of the ROC Area is near 1 and way above the non-discriminative characteristic (N.D) which represents equal TP and FP rates. The ROC curves for detection of not miner instances and miner instances are shown in Figure 6 and Figure 7 respectively.

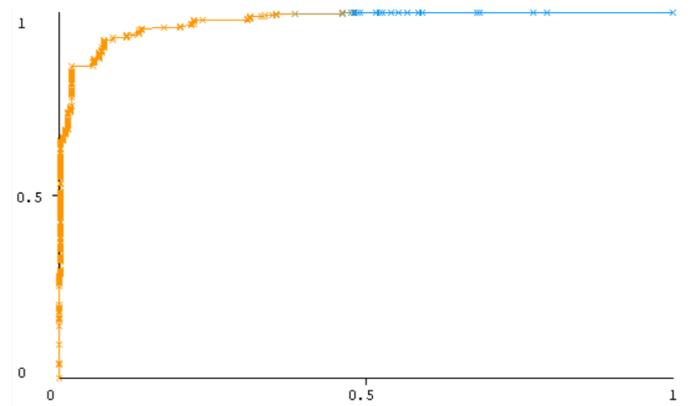

Figure 7. ROC curve for the "Not Miner" class

The ROC Area entails the predictive characteristics of the model to distinguish between the true positives and the true negatives. As such, the model does not only predict a positive value as a positive but as well as a negative value as a negative. The TP Rate represents the instances that are correctly classified as a given class which essentially is the rate of true positives. The FP Rate represents which of the instances falsely classified as a given class which essentially is the rate of false positives.

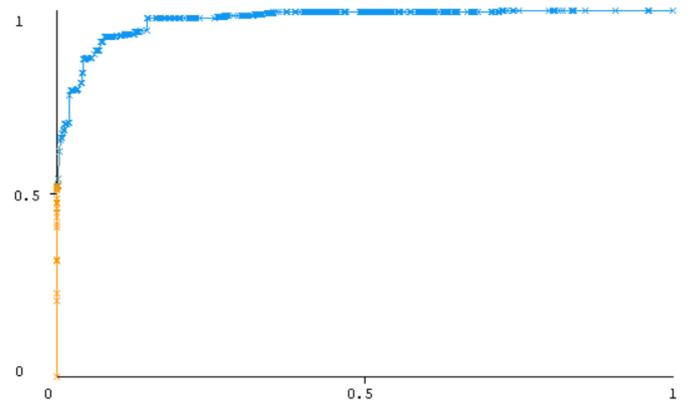

Figure 8. ROC curve for the "Miner" class

The PRC, as opposed to the ROC Area, represents the behavioral characteristics of Precision Vs Recall. The Precision value denotes the ratio of instances that are true of a given class divided by the sum of instances classified as that given class. The Recall value denotes the ratio of instances classified as a class divided by the actual sum in that given class. As such, this is equivalent to the TP rate. The F-Measure is a combined measure that depicts the ratio of double the product of Precision and Recall divided by the sum thereof, i.e. $\frac{2 \cdot Precision * Recall}{\sum(Precision + Recall)}$. The MCC is the measure of the quality of binary classifications taking into account true and false positives and negatives. It is a balanced measure that has a range [-1, 1], with -1 denoting a completely wrong classifier and 1 indicating the opposite. The variations in the clustering coefficient and vertex degree in these network traffic statistics in the respective clusters depict the overall movement of the movement vector from the feature centroid. Table V summarizes the differences between our prosed model and existing approaches.

TABLE V. COMPARISON WITH OTHER WORKS

| Attribute/ Model | Flexibility to Large-Scale Networks | Dynamic Complex Network modeling | Attack Network Formulation | Evaluation of Detection Model | Not Dependent on Attack Vector |
|---|---|---|---|---|---|
| Saad et. Al [20] | ✓ | ✗ | ✓ | ✓ | ✗ |
| Zareh et. al [21] | ✗ | ✗ | ✗ | ✓ | ✓ |
| Eskandari et. al [22] | ✓ | ✗ | ✗ | ✗ | ✗ |
| Carlin et. al [23] | ✓ | ✗ | ✗ | ✓ | ✗ |
| Veselý et. al [24] | ✓ | ✗ | ✗ | ✓ | ✓ |
| Musch et. al [29] | ✓ | ✗ | ✗ | ✓ | ✗ |
| Proposed Model | ✓ | ✓ | ✓ | ✓ | ✓ |

As can be seen in Table V, our modeling and detection approach has several advantages not limited to dependency on the prevailing attack vector (i.e. browser-based or installable binary-based) and incorporation of complex network modeling for effective detection.

V. CONCLUSIONS

The results presented in this paper demonstrate that the integration of semi-supervised learning with complex network theory modeling is effective at detecting crypto mining activities in a network environment. Our model's efficiency was enhanced by first clustering unlabeled data based on dynamic complex network characteristics and classifying the resultant clusters using a proximity-based classification algorithm, hence semi-supervised learning. The dynamic network characteristics exhibited in the network traffic generated by crypto mining activities serve as the modeling basis for detection. The presence of such crypto mining traffic in a corporate network is a high indicator of compromise. Our proposed detection methodology is advantageous in that it's independent of the nature of the victim device nor the underlying operating system since it's solely based on dynamic network statistics. Such an approach finds wide application in heterogeneous networks with varied devices such as IoT, SCADA/ICS systems, critical infrastructure, cloud computing, and so forth.